\documentclass[twocolumn,letterpaper,showpacs,floatfix,aps,amsfonts]{revtex4}

\usepackage[dvips]{graphics}

\def \beq {\begin{equation}}
\def \eeq {\end{equation}}

\begin{document}

\title{$N$-dimensional Vaidya metric with cosmological constant in double-null 
coordinates}

\author{Alberto Saa}
\email{asaa@ime.unicamp.br}
\affiliation{
Departamento de Matem\'atica Aplicada, \\
IMECC -- UNICAMP,
C.P. 6065, \\ 13083-859 Campinas, SP, Brazil.}

\pacs{04.50.+h, 04.20.Dw , 04.70.Bw ,   04.20.Jb}

\begin{abstract}
A recently proposed approach to the construction of the Vaidya metric in double-null
coordinates for generic mass functions is extended
to the $n$-dimensional $(n>2)$ case and to allow the inclusion
of a cosmological constant. 
The approach is based on a qualitative study of the null-geodesics, allowing the
description of   light-cones and revealing many features of the underlying causal
structure. Possible applications are illustrated by   explicit examples.  
Some new exact solutions are also presented and discussed.
The results presented here can simplify considerably 
the study
of spherically symmetric 
gravitational collapse and mass accretion in arbitrary dimensions. 
\end{abstract}

\maketitle

\section{Introduction}

The Vaidya metric\cite{Kramer} is a solution
of  Einstein's equations 
for a  spherically symmetric body with a  unidirectional radial null-fluid.
It has been used in the analysis of spherically symmetric
collapse and the
formation of naked singularities for many years (For references, see
the extensive list of \cite{Lake} and also \cite{Joshi}).
It is also  known that Vaidya metric can be obtained from the Tolman metric 
    by taking appropriate
limits in the self-similar case\cite{LemosHellaby}. This result has
shed some light on the nature of the so-called shell-focusing
singularities\cite{EardleySmarr}, as discussed in details
in \cite{Lake,Kuroda1,LakeZannias1,LakeZannias2}. The Vaidya metric has also proved to 
be useful in the study of Hawking radiation
and the process of black-hole 
evaporation\cite{Hiscock,Kuroda,Biernacki,Parentani},
in the stochastic gravity program\cite{Bei-Lok},  and,
more  recently, in the quasinormal modes analysis of varying mass black holes\cite{ACS,ACS2}.

The $n$-dimensional  Vaidya metric  is required 
in many physically relevant situations. The study of the gravitational collapse in higher
dimensional spacetimes\cite{GD}, for instance, has contributed to the elucidation of 
the formation, nature and   eventual visibility of singularities. This last topic
belongs to the realm of the celebrated Penrose's Cosmic Censorship Conjecture, see, {\em e.g.},
\cite{Joshi} for references. Higher dimensional varying mass black holes are also 
central protagonists
in the new phenomenological models with extra dimensions\cite{Arkani}. These objects
might be 
  obtained in the LHC at CERN\cite{LHC}  and, once produced, they
  are expected to decay driven by the emission of Hawing radiation. 
 The analysis of their quasinormal modes can help the understanding of the
 dynamics of the evaporation process, and could even lead to some 
 observational signs\cite{ACS2}.

The inclusion of a cosmological constant $\Lambda$ in the $n$-dimensional Vaidya metric
is mainly motivated   by the recent intensive 
activities in the AdS/CFT and dS/CFT conjectures, see \cite{conj} for some references. 
A cosmological constant
 is also necessary to allow the discussion of the $n=3$ case in the same framework
 of the higher dimensional cases.

The $n$-dimensional Vaidya metric was first discussed in \cite{IV}.
It can be easily cast  in $n$-dimensional 
($n>3$, the three dimensional case will be discussed latter) 
radiation coordinates
$(w,r,\theta_1,\dots,\theta_{n-2})$\cite{GD}:
\beq
\label{Vaidya}
ds^2 = -\left(1-\frac{2m(v)}{(n-3)r^{n-3}}\right)dv^2+2cdrdv + r^2d\Omega^2_{n-2},
\eeq
 where   $c=\pm 1$ and $d\Omega^2_{n-2}$ stands for the metric of the unity $(n-2)$-dimensional
sphere, assumed here to be spanned by the angular coordinates $(\theta_1,  \theta_{2},\dots ,\theta_{n-2})$,
\beq
d\Omega^2_{n-2} = \sum_{i=1}^{n-2}\left(\prod_{j=1}^{i-1} \sin^2\theta_j \right) d\theta_{i}^2.
\eeq
For the case of an ingoing radial flow, $c=1$ and
$m(v)$ is a monotone increasing mass function in the advanced
time $v$, while $c=-1$ corresponds to an outgoing radial flow,
with $m(v)$ being in this case 
a monotone decreasing mass function in the retarded
time $v$. The 4-dimensional Vaidya metric with a cosmological constant in
the radiation coordinates has  been   considered previously in \cite{PR}.

It is well   known   that the radiation coordinates
are defective at the horizon\cite{LSM}, implying that
the Vaidya metric (\ref{Vaidya}) is not geodesically complete   in any dimension. 
(See \cite{Fayos} for a discussion about possible   analytical
extensions). Besides, the cross  term $drdv$ 
can introduce unnecessary 
oddities in the hyperbolic equations governing 
the evolution of physical fields on spacetimes with the metric
  (\ref{Vaidya}). Typically, the double-null coordinates are   far more convenient.
This was the main motivation of  
 Waugh and Lake's work\cite{WL}, where  
the problem of casting the 4-dimensional Vaidya metric with $\Lambda=0$
in double-null coordinates
  is considered.
As all previous attempts to construct a general transformation from
radiation to double-null coordinates have failed, they followed 
Synge\cite{Synge} and considered Einstein's equation with
spherical symmetry in double-null coordinates {\em ab initio}.
The resulting equations, however, have revealed to be not analytical solvable for generic mass
functions.
  Waugh and Lake's work was recently revisited in \cite{GS}, where a semi-analytical approach
allowing for generic mass functions was proposed. Such approach 
consists in a qualitative study of the null-geodesics, allowing the
description of   light-cones and revealing many features of the underlying causal
structure. 
It can be used also for more quantitative analyses, indeed, it
has already enhanced considerably
the accuracy of the quasinormal modes analysis of 4-dimensional varying mass black holes \cite{ACS,ACS2}, 
and it can be also   applied to the study of gravitational collapse\cite{GS}.
We notice that another method to construct conformal diagrams based on a systematic
study of the null-geodesics was also recently proposed in \cite{meth}.

Here, we extend the approach proposed in \cite{GS} to the $n$-dimensional $(n>2)$ case and to 
allow the inclusion of a cosmological constant $\Lambda$. 
Some new exact solutions are also presented. For $n=3$, notably, a crucial non-linearity
can be circumvented and the problem can be reduced to the solution of a  second order linear
ordinary differential equation, opening the possibility for analytical 
study of large classes of mass functions.
We   generalize one of Waugh and Lake's exact solutions by introducing the case 
corresponding to $m(v)\propto v^{n-3}$, $n>3$, describing a naked or shell
focusing singularity in a $n$-dimensional spacetime.
We present also some explicit examples of the use of 
the semi-analytical approach to the study of the causal
structure of some particular solutions.

In the next Section,
the main equations are derived and the exact solutions are presented.
Section \ref{sec3} is devoted to the introduction of 
the semi-analytical approach. Some explicit examples are presented, particularly
the two apparent horizons case of an evaporating black hole in a de Sitter spacetime and
the BTZ black hole undergoing a finite time interval mass accretion process.
The last Section is left to some concluding remarks. This work has   two further appendices.
The first one presents explicitly the $n$-dimensional geometrical
quantities necessary to reproduce
the equations of the Section \ref{sec2}, and the last   contains the polynomial manipulations involved
in the calculations of the apparent horizons of Sections \ref{sec2} and \ref{sec3}.

\section{The metric}
\label{sec2}

The $n$-dimensional spherically symmetric line element in double-null coordinates $(u,v,\theta_1,\dots,\theta_{n-2})$ is
given by
\beq
\label{uv}
ds^2 = -2f(u,v)du\,dv + r^2(u,v)d\Omega^2_{n-2},
\eeq
where $f(u,v)$ and $r(u,v)$ are non vanishing smooth functions. We adopt here the same conventions
of \cite{WL} and \cite{GS}. In particular, the indices ${}_1$  and ${}_2$ stand for the
differentiation with respect to $u$ and $v$, respectively.

The energy-momentum tensor
of a unidirectional radial null-fluid  
in the eikonal approximation is 
given by
\beq
\label{T}
T_{ab} = \frac{1}{8\pi}h(u,v)k_a k_b,
\eeq
where $k_a$ is a radial null vector. 
We will consider here, without loss
of generality, the case of a flow along the $v$-direction. 
  The case of simultaneous ingoing and outgoing
flows in 4 dimensions was already been considered in \cite{Let}. Einstein's equations are less
constrained in such case, allowing the construction of
 some exact similarity solutions, which, incidentally, can
 also be generalized to the $n$-dimensional case in the light of the present work.
 
  Einstein's equations with  cosmological constant $\Lambda$
\beq
\label{eee}
R_{ab} - \frac{1}{2}g_{ab}R = -\Lambda g_{ab} + 8\pi T_{ab}
\eeq
implies that, for the 
for the energy-momentum tensor (\ref{T}),  
\beq
\label{contr}
R = \frac{2n}{n-2}\Lambda.
\eeq

Using  (\ref{ricci}) and (\ref{contr}),   Einstein's equations   for the metric (\ref{uv}) and the
energy-momentum tensor (\ref{T}) read 
\begin{eqnarray}
\label{eq1}
& & \frac{f_1}{f} - \frac{r_{11}}{r_1} = 0 \\
\label{eq2}
& & \frac{f_2}{f} - \frac{r_{22}}{r_2}  =  \frac{h}{n-2} \frac{r}{r_2} \\
\label{eq3}
& & \frac{f_1f_2}{f^2} - \frac{f_{12}}{f} - (n-2)\frac{r_{12}}{r} = -\frac{2\Lambda}{n-2} f\\
\label{eq4}
& & 2 \left(rr_{12} + (n-3)r_1r_2 \right) + (n-3)f = \frac{2\Lambda}{n-2}fr^2 \quad\quad
\end{eqnarray}
 For $n\ne 3$, differentiating Eq. (\ref{eq4}) with respect to $u$ and then
inserting Eq. (\ref{eq1}) leads to
\beq
\label{A}
\frac{r^{n-2}r_{12}}{f} - \frac{2\Lambda}{(n-2)(n-1)}r^{n-1} = -A,
\eeq
where $A(v)$ is an  arbitrary integration function.
The $n=3$ case will be considered below. Now, differentiating Eq. (\ref{eq4}) with respect to $v$ and
using (\ref{eq2}) and (\ref{A})   gives
\beq
\label{hh}
h = -\left(\frac{n-2}{n-3}\right)  \frac{fA_2}{r^{n-2}r_1}.
\eeq

Eq. (\ref{eq1}) is ready to be integrated  
\beq
\label{B}
f = 2Br_1,
\eeq
where $B(v)$ is another arbitrary integration (and nonvanishing) function. 
From   (\ref{hh}) and (\ref{B}), one has
\beq
\label{h}
h = -2\left(\frac{n-2}{n-3}\right)B \frac{A_2}{r^{n-2}}.
\eeq
Finally, by using (\ref{eq4}) and (\ref{B}), Eq. (\ref{A}) can be written as
\beq
\label{r2}
r_2 = -B \left(1 - \frac{2A}{(n-3)r^{n-3}} - \frac{2\Lambda}{(n-2)(n-1)}r^2 \right).
\eeq
Note that (\ref{eq3}) follows from Eqs. (\ref{B}) and (\ref{r2}). Einstein's equations 
are, therefore, equivalent to the equations (\ref{B}), (\ref{h}), and (\ref{r2}),
generalizing the results of \cite{WL} and \cite{GS}.

In order to  interpret physically  the arbitrary integration functions $A(v)$ and $B(v)$,
let us transform from the double-null coordinates back to the radiation coordinates by
means of the coordinate  change $(u,v)\rightarrow (r(u,v),v)$. Such  a transformation casts
(\ref{uv}) in the form
\beq
\label{uv1}
ds^2 = 4Br_2 dv^2 - 4Bdrdv + r^2d\Omega^2_{n-2},
\eeq
where (\ref{B}) was explicitly used. Comparing (\ref{Vaidya}) and (\ref{uv1}) and taking into account
(\ref{r2}), it is clear that with the choice
\beq
\label{cC}
B = -\frac{1}{2}\frac{A_2}{|A_2|},
\eeq
for $A_2\ne 0$, 
the function $A(v)$ should represent the mass of the $n$-dimensional
solution, 
  suggesting the following $n$-dimensional generalization of the mass definition 
  based in the angular
components of the Riemann tensor proposed for for $\Lambda=0$ in \cite{mass}:
\beq
\label{massd}
m = \frac{n-3}{2}r^{n-3} R_{\theta_1\!\theta_k\!\theta_1 }^{\phantom{\theta_j\!\theta_k\!\theta_j\!}\theta_k}, 
\eeq
$k>1$, see   (\ref{m}) and (\ref{r2}). 
Note that for constant $A$, the choice of the 
function $B(v)$ is irrelevant  since it can be absorbed by a redefinition of $v$.
Note also that the weak energy condition applied for (\ref{T}) requires, from (\ref{h}),
that $BA_2 \le 0$. Since $B$ must be nonvanishing  from (\ref{B}), $A(v)$ must be
a monotone function.

The Kretschmann scalar $K=R_{abcd}R^{abcd}$ could be also invoked to interpret the 
integration function $A$. Taking into account
Eqs. (\ref{B}) and (\ref{r2}), we have from (\ref{K})
\beq
\label{KN}
K = 4\frac{(n-1)(n-2)^2}{(n-3)}\frac{A^2}{r^{2(n-1)}} + \frac{8n}{(n-1)(n-2)^2}\Lambda^2.
\eeq
It is clear from (\ref{KN}) that $r=0$ is a singularity for $A\ne 0$ and that $A$ acts as a gravitational
source 
  placed at $r=0$.

The problem now may be stated in the same way of the 4-dimensional $\Lambda=0$
case\cite{GS}: given the
mass function $A(v)$ and the constant $B$, 
one needs to solve
Eq. (\ref{r2}), giving rise to the function $r(u,v)$.
  Then,  
$f(u,v)$ and $h(u,v)$ are calculated from (\ref{B}) and
(\ref{h}). The arbitrary function  of $u$ 
appearing in the integration of (\ref{r2}) must be chosen properly\cite{WL} in order to have
a non-vanishing $f(u,v)$ function from (\ref{B}). 
Unfortunately,
as stressed early by Waugh and Lake\cite{WL},
such a procedure is not analytically
solvable in general. They,
  nevertheless, were able to find some regular solutions for
Eqs. (\ref{B})-(\ref{r2}) for $n=4$ and $\Lambda=0$, namely the  linear ($A(v)=\lambda c v$)
and a certain exponential ($A(v)=\frac{1}{\beta}\left(
\alpha\exp(\beta c v/2) + 1 \right) $) 
mass functions ($\lambda,\alpha$, and $\beta$ are 
positive constants, $c=\pm 1$, corresponding to ingoing/outgoing
flow, respectively). 
The 4-dimensional linear mass case was also considered in \cite{LakeZannias2} 
in great detail and in a more general situation (the case of a
charged radial null fluid).
In \cite{GS}, another 4-dimensional exact solution corresponding to
$A(v) = \kappa/v$  and $\Lambda=0$ was also presented and discussed.
These are the only 
varying mass analytical solutions obtained in 
double-null coordinates so far. We notice, however, that Kuroda 
was able to construct a transformation from radiation to double-null
coordinates for some other particular mass functions in four dimensions\cite{Kuroda}.

In the following Section, we will present a semi-analytical procedure
to attack the problem of solving Eqs. (\ref{B})-(\ref{r2}) for 
general mass functions obeying the weak energy condition,
generalizing in this way the results of \cite{GS} obtained for $n=4$ and $\Lambda=0$.
The approach allows us to construct qualitatively conformal diagrams,
identifying horizons and singularities, 
and also to evaluate specific geometric quantities.
Before, however, we notice that the Waugh and Lake's linear 4-dimensional solution
can be also generalized for $n$-dimensions. Let us consider the mass function
\beq
\label{massg}
A(v) = \lambda v^{n-3},
\eeq
$v\ge 0$, $\lambda > 0$.
For this mass function and $\Lambda=0$, with the choice (\ref{cC}), Eq. (\ref{r2}) reads
\beq
\label{eq21}
r_2 = \frac{1}{2}  - \frac{\lambda}{n-3}\left( \frac{v}{r}\right)^{n-3},  
\eeq
which can be integrated as
\beq
\label{ada}
\ln v + \int^{(r/v)}  \left(s - \frac{1}{2} + \frac{\lambda}{n-3}s^{-(n-3)} \right)^{-1} ds = D(u),
\eeq
where $D(u)$ is an arbitrary integration function, implying, from Eq. (\ref{B}),
that
\beq
\label{ff}
f = -\frac{v^{n-2}D_1}{r^{n-3}}\left[ \left(\frac{r}{v} \right)^{n-2} - \frac{1}{2}\left(\frac{r}{v} \right)^{n-3} + \frac{\lambda}{n-3}\right].
\eeq
The integration function $D$ must be chosen in order to have a regular $f$.
As in the 4-dimensional case originally
considered by Waugh and Lake, we have three qualitatively distinct cases:
 $0<\lambda < \lambda^{c}_n$,  $\lambda = \lambda^{c}_n$, and $\lambda > \lambda^{c}_n$, 
 where
\beq
\label{lambdac}
\lambda^{c}_n = \left(\frac{n-3}{2(n-2)} \right)^{n-2}.
\eeq
For $\lambda > \lambda^{c}_n$, for instance, 
the quantity inside the square brackets in (\ref{ff}) 
does not vanish, see   Appendix \ref{hor}.
The integral in (\ref{ada}) can be (numerically) evaluated and, for instance with the 
Waugh and Lake's choice $D(u) = -u$, $r(u,v)$ can be also (numerically) determined.
The resulting causal structures are the same ones as the Waugh and Lake's 4-dimensional cases.
We will construct the relevant conformal diagrams 
for the three qualitatively distinct cases
as an application of the semi-analytical
approach. 
\subsection{The three dimensional case}

In lower dimensional  ($n<4$) spacetimes, the Riemann tensor is completely determined by
its traces, namely the Ricci tensor and the scalar curvature, implying
some qualitative distinct behavior for the solutions of Einstein's
equations in these situations\cite{lower}.
For $n=3$, Eq. (\ref{eq4}) reads
\beq
\label{eq4-3}
\frac{rr_{12}}{f} + {\Lambda}r^2 = 0.
\eeq
It is interesting to compare this last equation with (\ref{A}).
Eq. (\ref{B}) is still valid  and by using it, Eq. (\ref{eq4-3}) can be easily integrated,
\beq
  r_2 + {\Lambda}Br^2   = C,
\eeq
where $C$ is an arbitrary integration function   which we call, by convenience, $C(v) = -B(v)A(v)$,
leading to
\beq
\label{r2-3}
r_2 = -B \left( -A - \Lambda r^2 \right),
\eeq
which
corresponds to the 3-dimensional counterpart of Eq. (\ref{r2}).    Eq. (\ref{eq3}) for $n=3$ also
follows from (\ref{B}) and (\ref{r2-3}). As for the $n$-dimensional case, by using (\ref{B}) and (\ref{r2-3}),
one gets from (\ref{eq2}) the last equation
\beq
\label{h-3}
h = -B \frac{A_2}{r}.
\eeq
One can show that
the integration functions $A$ and $B$ have the same physical interpretation as the 
higher dimensional cases  by introducing the radiation coordinates. We have
for the 3-dimensional case
\beq
ds^2 = -(2B)^2\left(  - A - \Lambda r^2 \right) dv^2 - 4Bdrdv + r^2d\theta^2.
\eeq
It is clear that with the choice (\ref{cC}) for $B$, the function $A$ plays the role of the
BTZ black-hole mass\cite{BTZ} ($\Lambda<0$ in this case). Note, however, that for $n=3$ there is no 
purely angular components of the Riemann tensor and, therefore, there is no equivalent of
the mass definition (\ref{massd}). The 
Kretschmann scalar in this case reads simply
\beq
K = 12\Lambda^2.
\eeq
In fact, if terms involving distributions are take into account properly,   curvature invariants  as $K$ reveal
that the origin for the BTZ black hole is a conical singularity (See, for a recent rigorous analysis, \cite{distr}).

In   contrast with the $n$-dimensional case, the non-linearity present in (\ref{r2-3})
can be easily circumvented. Let us consider $A_2>0$, $B=-1/2$, and $\Lambda = -1/\ell^2$ (Other cases
follow straightforwardly). By introducing the linear second order  ordinary differential equation
\beq
\label{ode}
w''(v) - \frac{A(v)}{4\ell^2}w(v) = 0,
\eeq
it is easy to show that 
\beq
r(u,v) = -2\ell^2 \frac{P(u)w'_a(v) + w'_b(v)}{P(u)w_a(v) + w_b(v)} ,
\eeq
is   solution of (\ref{r2-3}),
where $w_a(v)$ and $w_b(v)$ are the two linearly independent solutions of (\ref{ode}),
and $P(u)$ is an arbitrary function.  The second order linear equation
(\ref{ode}) has analytical solution in closed form for many functions $A(v)$. For 
instance, for $A(v)\propto v^\alpha$   (or $A(v) \propto \exp\alpha v$), $\alpha \in \mathbb{R}$, Eq. (\ref{ode})
is equivalent to the Bessel equation, after an appropriate redefinition of $v$
($v\rightarrow v^{(\alpha/2+1)}$ or $v\rightarrow e^{\alpha v/2}$).
Even for the case where no solution in closed form can be obtained, 
the main analytical properties of $w(v)$ can be inferred easily from   (\ref{ode}) due to
its linearity.

\section{Semi-analytical approach}
\label{sec3}

We review here the semi-analytical approach proposed in \cite{GS} for the solution of the
  equation  
  (\ref{r2}) (or, analogously, (\ref{r2-3})  for the 3-dimensional
case). Firstly, notice that in double-null coordinates the light-cones correspond to the
hypersurfaces with constant $u$ or constant $v$. One can, in this case, deduce the causal
structure of a given spacetime by considering the set of null geodesics corresponding to  
$u=$ constant or $v=$ constant. The function $r(u,v)$, obtained as the solution of (\ref{r2}) 
  with a careful choice  of   initial conditions\cite{WL,GS}, 
has also a clear geometrical interpretation: it is
the radius of the $(n-2)$ dimensional sphere defined by the intersection between the hypersurfaces 
$u=$ constant and $v=$ constant. 
Once $r(u,v)$ is obtained from (\ref{r2}), $f(u,v)$ and $h(u,v)$ can be calculated directly from 
Eqs. (\ref{B}) and (\ref{h}), giving all the information about the spacetime in question.
Incidentally, Eqs. (\ref{r2}) along constant $u$, and, consequently, along a portion of the light-cone, is
a first order ordinary differential
equation in $v$. One can evaluate the function $r(u,v)$ in
any spacetime point by solving the $v$-initial value problem knowing
$r(u,0)$. For instance, the usual constant curvature empty spacetimes ($A=0$) can be obtained 
by  choosing $r(u,0)=u/2$, leading to
\beq
\label{dS}
  \int_{r_0}^r \left(1-\frac{2\Lambda}{(n-2)(n-1)}s^2\right)^{-1} ds = \frac{1}{2}(v+ u). 
 \eeq
 As it is expected, 
 for the $n$-dimensional de Sitter case ($\Lambda > 0$),
  $r =  \sqrt{ (n-2)(n-1)/2\Lambda}$  corresponds to a 
  cosmological horizon. The condition $r_1(u,0)\ne 0$ is sufficient to assure that $f(u,v) \ne 0$ everywhere\cite{GS}.

\begin{figure}[ht]
\resizebox{\linewidth}{!}{\includegraphics*{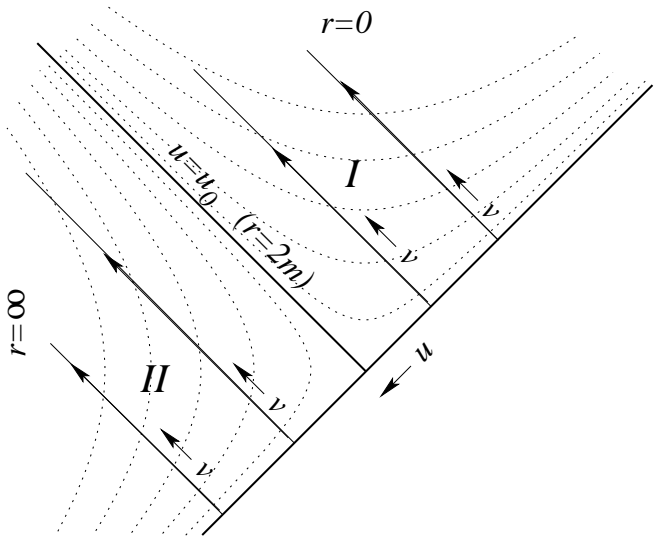}}
\caption{
The solutions $r(u,v)$ of (\ref{r2}) for the Schwarzschild case ($A(v) = m$,
see \cite{WL}). The dotted lines correspond to the lines of constant $r$. In the
region I (the interior region), the values of $r$ decreases monotonically upward,
varying from $r=2m$, the horizon corresponding to the degenerated hyperbole 
$u=u_0$ and $v=v_0$, to the singularity $r=0$. In the exterior region (II), on the
other hand, the values of $r$ increases monotonically leftward, varying from $r=2m$
to the spatial infinity $r=\infty$. Some slices of constant $u=\bar{u}$ are depicted by the
lines with arrows. The values of $r(\bar{u},v)$ read along such lines correspond to the solutions
of (\ref{r2}) with a certain initial condition $r(0)=r_{\bar{u}}$. Since the lines of constant
$u$ are portions of the light cone, the function $r(\bar{u},v)$ describe how the light cones cross 
the $(n-2)$-dimensional spheres of radius $r$. In the interior region, for instance, all   null geodesics
eventually
reach the singularity, whereas in the exterior, the null geodesics with constant $u$
escape to the null infinity. Between such regions, there is an event horizon at $u=u_0$. We notice
that all this
information is coded in the qualitative behavior of the solutions $r(\bar{u},v)$ for 
the first order ordinary differential equation
corresponding to (\ref{r2}) considered along constant $u=\bar{u}$. This is the basic idea of
our semi-analytical approach.
}
\label{schem}
\end{figure}

Let us focus now on the solutions $r(v) = r(\bar{u},v)$ of (\ref{r2}), considered as a 
a first order ordinary differential
equation in $v$ for constant $u=\bar{u}$, keeping in mind that they indeed describe how $r(\bar{u},v)$ 
varies along the $u=\bar{u}$   portion of the light-cone and, thus,
that they   are closely related to the causal 
structure of the underlying spacetime.
This situation is shown schematically in Fig. \ref{schem}.

\begin{figure}[ht]
\resizebox{\linewidth}{!}{\includegraphics*{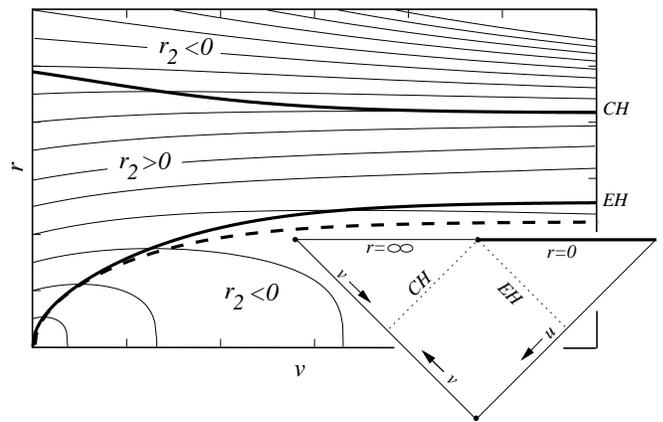}}
\caption{In the region below the   curve 
$r_{EH}(v)$ (the apparent event horizon),  
  $r_2<0$. Any solution $r(v)$ of (\ref{r2}) that enters into this region will reach
the singularity at $r=0$ with finite $v$. Solutions
confined to the $r_2>0$ region  always escape from the singularity and
tend asymptotically to $r_{CH}(v)$ (the apparent cosmological horizon). 
The region  above $r_{CH}(v)$ corresponds to the cosmological exterior,
any solution there also tends asymptotically to $r_{CH}(v)$.
The dashed line would correspond to the apparent event horizon for the case of a 
vanishing $\Lambda$
(The unique solution of (\ref{ahor}) for $\Lambda=0$).
The conformal diagram is inserted. 
(The case depicted here corresponds to: $n=5$, $\Lambda=1$, and 
  $A(v) = (6/5)\tanh v$, $v\ge 0$. In this work, we do not consider maximal extensions.
  For a discussion  of maximal conformal diagram for de Sitter and anti-de Sitter black
  holes, see \cite{BH}, for instance.
  ) }
\label{figure1}
\end{figure}

For $n>3$, the curves  defined by
\beq
\label{ahor}
-\frac{2\Lambda}{(n-2)(n-1)} {r}^{n-1} +  {r}^{n-3} = \frac{2A(v)}{n-3}
\eeq
correspond (if they indeed exist) to the frontiers of certain regions of the
$(v,r)$ plane where the solutions $r(v)$ of (\ref{r2}) have qualitative
distinct behaviors. 
For the de Sitter case, (\ref{ahor}) can define  two non intersecting curves,
whereas for the anti de Sitter case $(\Lambda < 0)$ there is only one curve. (See Appendix \ref{hor}).
Let us suppose, for instance, a  de Sitter case with $A_2(v)>0$ (and
$B=-1/2$. Outgoing radiation flows, 3-dimensional spacetimes, or the anti de Sitter case follow
in a straightforward manner).  
Let us call $r_{EH}(v)$ and $r_{CH}(v)$ the curves defined by (\ref{ahor}), see Fig. \ref{figure1}.
Due to (\ref{r2}), 
for all points of the plane $(v,r)$ below $r_{EH}(v)$,
 $r_2 < 0$. Hence, any solution $r(v)$ entering in this region
will, unavoidably, reach the singularity at $r=0$, with finite
$v$. 
Suppose a given solution $r_{\rm i}(v)$ with initial condition
$r_{\rm i}(0)=r_{\rm i}$ enters into the region below $r_{EH}(v)$. As, for
smooth $A$, the uniqueness of
solutions for (\ref{r2}) is guaranteed for any point with $r> 0$,
any solution starting at $r(0)<r_{\rm i}$ is confined the the
region below $r_{\rm i}(v)$  and will also reach the singularity
at $r=0$, with finite $v$. On the other hand, suppose that
a given solution $r_{\rm e}(v)$ with initial condition
$r_{\rm e}(0)=r_{\rm e}$ never enters into the region bellow the curve
$r_{EH}(v)$. 
Solution starting at $r(0)>r_{\rm e}$ are, therefore, confined to
the region above $r_{\rm e}(v)$ and will escape from the
singularity at $r=0$. Hence, one has   two qualitatively distinct behavior
for the light-cones, see Fig. \ref{figure1}. For $r(0)<r_{\rm i}$, the future direction
points toward the singularity and   all null geodesics eventually reach
$r=0$. This is the typical situation in the interior region of a black-hole.
For $r(0)>r_{\rm e}$, the $u=$ constant portion of the light cones escape from
the singularity.
 The curve  $r_{EH}(v)$ plays
the role of an apparent event horizon.
For a given $v$, all 
solutions $r(v)$ in Fig. \ref{figure1} such that $r(v)\le r_{EH}(v)$ will be captured
by the singularity. Solutions for which $r(v)> r_{EH}(v)$ are temporally
free, but they may find themselves trapped later if $r_{EH}(v)$ increases.
The event horizon   corresponds to the last of these solutions trapped by the singularity  and
  is located somewhere between $r_{\rm i}(v)$ and $r_{\rm e}(v)$.

In the region between $r_{EH}(v)$ and $r_{CH}(v)$, the solutions $r(v)$
have $r_2 > 0$. They escape from $r_{EH}(v)$ and tend asymptotically to $r_{CH}(v)$,
which plays the role of an apparent cosmological horizon. The region above
$r_{CH}(v)$ corresponds to the cosmological exterior (See, for instance, \cite{Carter}. For $A=0$, 
the cosmological exterior region 
corresponds to 
  the choice $r_{0}=\infty$ in (\ref{dS})). There, $r_2<0$ and the solutions 
also tend asymptotically to $r_{CH}(v)$.
The case where $r_{EH}(v)$ and $r_{CH}(v)$ converge to the same function gives rise
to the interesting $n$-dimensional Nariai solution\cite{CDL}.

Summarizing, since the partial differential equation (\ref{r2}) for $r(u,v)$  
does not involve $u$ explicitly, it is always possible to  discover how the light-cones cross the $(n-2)$-dimensional
spheres
of radius $r$ by exploring, in the plane $(v,r)$, the solutions $r(\bar{u},v)$ of the first order 
  differential
equation corresponding to (\ref{r2}) considered on the constant $u$ null-geodesics.
From these informations, one can
reconstruct the causal structure of the underlying spacetime. For   typical cases,
 we will have some spacetime regions where the light cones necessarily end  in the
singularity at $r=0$, whereas for some other regions it is always possible to escape to the null
infinity. The frontier of these regions must correspond  to event horizons.
 Let us illustrate the approach with  some  explicit examples.
 
 \subsection{Waugh and Lake generalized solutions}
 
 As the first application of the semi-analytical approach, let us consider the generalized
 Waugh and Lake solutions corresponding to the choice (\ref{massg}) for the mass
 function, $n>3$ , and $\Lambda =0$. From the discussion of Section \ref{sec2}, we expect
 three qualitatively distinct case according to the value of $\lambda$.
 In fact, these solutions have the same causal structure for any $n>3$.
 The frontier of the region in the $(v,r)$ plane where all the solutions of (\ref{eq21})
 reach the singularity at $r=0$   corresponds to the straight line
 \beq
 r_0(v) = \left(\frac{2\lambda}{n-3} \right)^{\frac{1}{n-3}} v.
 \eeq
 Taking the $v$-derivative of (\ref{eq21}), one gets
 \beq
 r_{22} =  \lambda\frac{v^{n-2}}{r^{n-3}}\left(
 -1 + \frac{1}{2}\frac{v}{r} -
 \frac{\lambda}{n-3}\left(\frac{v}{r}\right)^{n-3} \right).
 \eeq
 The regions in the plane $(v,r)$ where the solutions obeys
 $r_{22}=0$ are the straight lines defined by
 \beq
 \label{rr}
 \left(\frac{r}{v} \right)^{n-2} - \frac{1}{2}\left(\frac{r}{v} \right)^{n-3} + \frac{\lambda}{n-3}=0.
 \eeq
 One can now repeat the analysis done for the $n=4$ case by
 considering   the
 three qualitative different cases according to the value
 of $\lambda$ and the possible solutions of (\ref{rr}).
 For this purpose, we will consider the solutions of
 (\ref{eq21})
 with the initial condition $r(u,0)\propto u$.

\begin{figure}[ht]
 \resizebox{\linewidth}{!}{\includegraphics*{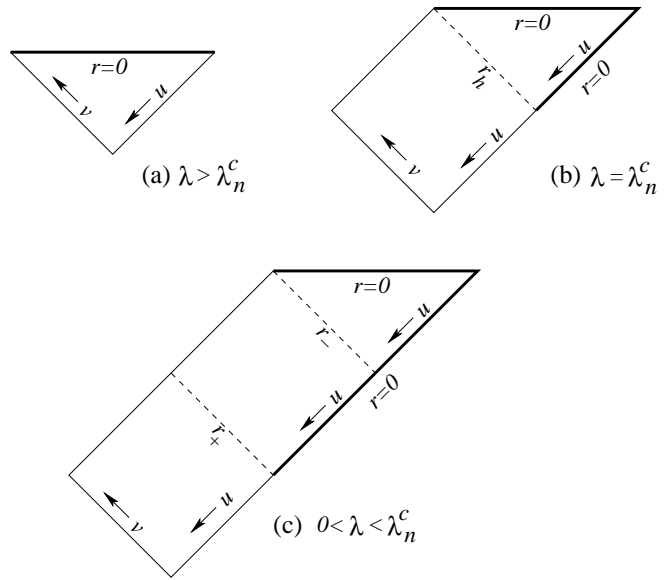}}
 \caption{Conformal diagrams corresponding to the mass function
 (\ref{massg}) and $\Lambda=0$. Cases (b) and (c) exhibit naked and shell focusing singularities.
 For all cases, 
 the curves in the $(v,r)$ are similar to those ones presented in
 \cite{GS}, corresponding to the $n=4$ case.}
 \label{figure2}
\end{figure}

 For $\lambda>\lambda^c_n$, with $\lambda^c_n$ given by (\ref{lambdac}),
 Eq. (\ref{rr}) has no solution and
 $r_{22}<0$ for all points with $v>0$. The only
 relevant frontier in the plane $(v,r)$ 
 is the $r_0(v)$ straight line. All solutions
 of (\ref{eq21}) are concave functions and cross the $r_2=0$ 
 line, reaching the singularity with a finite $v$ which increases
 monotonically with $u$.
 The causal structure of the corresponding spacetime is very
 simple. There is no horizon, and all future cones  
 end  in the   singularity at $r=0$. The associated conformal diagram is
 depicted in Fig. \ref{figure2}.

 For $\lambda=\lambda^c_n$, $r_{22}=0$ only on the line 
 \beq
 r_h(v) = \frac{n-3}{2(n-2)}v. 
 \eeq
 This straight line
 itself is a solution of (\ref{eq21}). All other solutions are concave functions.
 Note that, in this case, $r_h(v) > r_0(v)$ for $v>0$.
 We have two distinct qualitative behavior
 for the null trajectories along $u$ constant. All solutions
 starting at $r(0)>0$ are confined to the region
 above $r_h(v)$. They never reach the singularity, all
 trajectories reach $\cal I^+$. However, in the region below $r_h(v)$, 
 we have infinitely many concave trajectories starting and ending
 in the (shell-focusing) singularity. They 
  start at $r(0)=0$,
 increase in the region between  $r_h(v)$ and $r_0(v)$, cross
 the last line and reach unavoidably $r=0$ again, with finite $v$.
 The trajectory $r_h(v)$ plays the role of an event horizon, separating
 two regions with distinct qualitative behavior: one where
 constant $u$ null trajectories reach $\cal I^+$ and another
 where they start and end in the singularity. 
 This behavior is only possible,
 of course, because the solutions of (\ref{eq21}) fail to be unique
 at $r=0$. The relevant conformal diagram is also shown in Fig. \ref{figure2}.

 For $0<\lambda<\lambda^c_n$, we have three distinct regions
 according to the concavity of the solutions. They are limited by the
 two straight lines $r_-(v)$ and $r_+(v)$ defined by (\ref{rr}). 
 We have $r_0(v) < r_-(v) < r_+(v)$ for $v>0$.
 Both straight lines $r_-(v)$ and $r_+(v)$
   are also solutions
 of (\ref{eq21}). Between them, solutions are
 convex. Above $r_+$ and below $r_-$, solutions are concave. The last
 line is the inner horizon. Inside, the null trajectories with constant $u$
 start and end in the singularity $r=0$. The line $r_+$ is an outer
 horizon. Between them, the constant $u$ null trajectories starts
 in the naked singularity and reach $\cal I^+$. Beyond the outer 
 horizon, all  constant $u$ null trajectories escape the singularity, see
 Fig. (\ref{figure2}).

\subsection{Black-hole evaporation in a de Sitter spacetime}
 
 As an example of using the semi-analytical approach for $\Lambda\ne 0$,
 let us consider the case of a evaporating black hole  due to Hawking
 radiation in a de Sitter $n$-dimensional ($n>3$) spacetime. 
 The mass decreasing rate for these $n$-dimensional black holes can
 be obtained  from the $n$-dimensional Stefan-Boltzmann law (see, for instance, \cite{CDM}), 
 leading to
\begin{equation}  
 \frac{dm}{du} = - {a}m^{-2/(n-3)}
  \label{dMdt}
\end{equation}
where $a>0$ is the effective $n$-dimensional Stefan-Boltzmann constant.
Eq. (\ref{dMdt}) can be immediately integrated, leading to 
\begin{equation}
\label{meva}
  m(u) = m_0\left(1 -   \frac{u}{u_{0}}\right)^{\frac{n-3}{n-1}},
  \label{real_mass}
\end{equation}
where $u_0$ in the lifetime  of a black hole with initial mass $m_0$,  
\beq
a u_0 = \frac{n-3}{n-1}m_0^{(n-1)/(n-3)}.
\eeq
We recall that (\ref{dMdt}) is not expected to be valid in the very
final stages of the black hole evaporation, where 
the appearance of new emission channels for Hawking radiation\cite{final} can induce
changes in the value of the constant $a$, and maybe even 
the usual adiabatic
derivation of Hawking radiations is  not valid any longer. We do
not address these points here. We assume that the black hole evaporates
completely   (\ref{meva})  for
all $0\le u \le u_0$ and that $m=0$ for $u>u_0$. Notice that, since $m'\rightarrow\infty$
for $u\rightarrow u_0$, a naked singularity will be formed after the vanishing of
the black hole\cite{Kuroda}.

\begin{figure}[ht]
 \resizebox{\linewidth}{!}{\includegraphics*{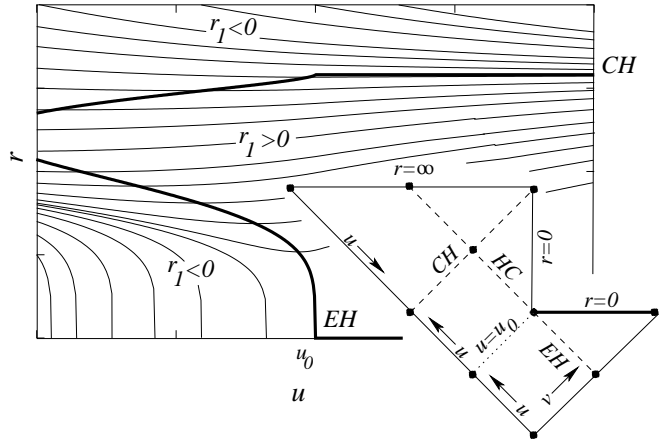}}
 \caption{The $(u,r)$ plane for an evaporating black hole in a $n$-dimensional 
 de Sitter spacetime. The apparent horizons $r_{EH}(u)$ and $r_{CH}(u)$ delimit
 the regions where the solutions of (\ref{eqq}) have qualitatively distinct 
 behavior. Since the black hole vanishes for $u=u_0$, for $u>u_0$ the only
 remaining horizon corresponds to $r_{CH}(u)$. The conformal diagram is inserted.
 For $u<u_0$, it represents an usual black hole. For $u>u_0$, we have the empty
 space left after the vanishing of the black hole. The dashed line $HC$ (extension
 of the black hole horizon $EH$) is a Cauchy horizon. Beyond it, we have a breakdown
 of predictability due to the naked singularity left by the black hole. Part of the cosmological
 exterior region (the region above $CH$ and below $HC$) is completely insensitive to the singularity. 
 (The case depicted here corresponds to: $n=6$, $\Lambda=5/2$, and 
  $m_0=u_0=1$, $u\ge 0$.) }
 \label{fig4}
 \end{figure}

We can consider now a $n$-dimensional Vaidya metric with a mass term of the
form given by (\ref{meva}). This is the first case of an outgoing radial
flux we consider,
and according to (\ref{cC}), $B=1/2$. 
However, in this case, $f(u,v)$ has different sign and the temporal directions
for the $B=-1/2$ case are now spacelike. The transformation $(u,v)\rightarrow (v, -u)$
restores the temporal and spatial directions\cite{GS}. One needs to keep in mind that, for
$B<0$ (ingoing radiation fields), $v$ corresponds to the advanced time, whereas
for $B>0$ (outgoing radiation fields), $u$ should correspond to the retarded 
time. In this case, the equivalent of 
Eq. (\ref{r2}) for the mass function   (\ref{meva}) reads
\beq
\label{eqq}
r_1 = \frac{1}{2} \left( 1 - \frac{2m_0}{n-3}\frac{\left(1-u/u_0\right)^{\frac{n-3}{n-1}}}{r^{n-3}} 
- \frac{2\Lambda}{(n-1)(n-2)}r^2 \right),
\eeq
$0\le  u\le 0$. For $u>u_0$, we have a empty $n$-dimensional de Sitter spacetime. 
We can now apply the same procedure we have used for the previous cases.
We call $r_{EH}(u)$ and $r_{CH}(u)$ the curves on the $(u,r)$ plane
delimiting the regions where the solutions of (\ref{eqq}) have qualitative
distinct behavior. (See Fig. \ref{fig4}).
 In this case, since $A_1<0$, the curve $r_{EH}(v)$ decreases and reach $r=0$
 for $u=u_0$, corresponding to the total vanishing of the black hole.
 After this point, $r_{CH}(v)$ is the only horizon (a cosmological one) of this
 spacetime. Before $u=u_0$, however, some solutions of (\ref{eqq}) that crossed
 $r_{EH}(v)$ could reach the singularity at $r=0$. The exterior cosmological region
 (region of the $(u,r)$ plane above $r_{CH}(v)$) is insensitive to the singularity.
 The resulting conformal diagram shown in Fig. \ref{fig4}. It represents a spacetime
 with a cosmological horizon and, in its interior, a black hole that evaporates
 completely leaving behind a naked singularity.

\subsection{Mass accretion by BTZ black-holes}
 
\begin{figure}[ht]
\resizebox{\linewidth}{!}{\includegraphics*{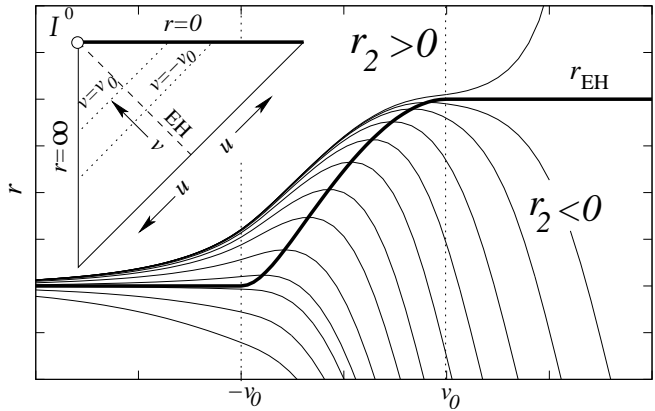}}
\caption{The $(v,r)$ plane for a BTZ black hole accreting mass according to (\ref{ac3}).
In the region below the   curve 
$r_{EH}(v)$ (the apparent event horizon),  
  $r_2<0$. The constant $u=\bar{u}$ 
  null geodesics associated to the solutions $r(\bar{u},v)$ of (\ref{r2-3}) entering into this region   reach
the conical singularity at $r=0$ with finite $v$. The null geodesics
confined to the $r_2>0$ region  can escape from the singularity, reaching the null and spatial
  infinity. 
The conformal diagram is inserted. 
(The case depicted here corresponds to:   $\Lambda=-1$, $B=-1/2$,  and 
  $m_{\rm i} =  m_{\rm f}/9 = 1$.  ) 
}
\label{btz}
\end{figure}

As an explicit example of the $n=3$ case, let us consider a BTZ black hole\cite{BTZ} undergoing 
a mass accretion process during a finite time interval. The $C^1$-class mass function
\beq
\label{ac3}
A(v) = \left\{ 
 \begin{array}{l}
  m_{\rm i},   \quad   v < -v_0,   \\
    av\left(3v_0^2-v^2\right) + b, \quad -v_0 \le v \le v_0, \\
  m_{\rm f},   \quad   v >  v_0.
  \end{array} 
\right.
\eeq
does correspond to such a situation. Here, $m_{\rm i}$ and $m_{\rm f}$ stand for, respectively,
  the initial and final mass, and the constants  $a = (m_{\rm f}-m_{\rm i})/4v_0^3$ and 
$b = (m_{\rm f}+m_{\rm i})/2$ are chosen in order to have
smooth matches at $v=\pm v_0$. For the BTZ case, $\Lambda = -1/\ell^2$ and the apparent horizon  
corresponding to the region of the $(r,v)$ plane where $r_2=0$
  is given simply by
$r_{\rm EH}(v)  = \ell \sqrt{A(v)}$, see (\ref{r2-3}). 
Since $A_2>0$, we adopt $B=-1/2$.
The solutions of (\ref{r2-3}) 
below the curve $r_{\rm EH}(v)$ have $r_2 <0$, whereas those ones above have $r_2 >0$, see Fig. \ref{btz}.
For $v<-v_0$, and analogously for $v>v_0$, Eq. (\ref{r2-3}) can be easily integrated
\beq
\label{ww}
\frac{r(u,v)-\ell\sqrt{m_{\rm i}}}{r(u,v)+\ell\sqrt{m_{\rm i}}} = D_{\rm i}(u)e^{ \sqrt{m_{\rm i}} v/\ell},
\eeq
where $D_{\rm i}(u)$ is a monotone arbitrary integration function. The region $r>\ell\sqrt{m_{\rm i}}$,
is obtained by choosing
$D_{\rm i}(u)>0$, whereas  $r<\ell\sqrt{m_{\rm i}}$ corresponds to $D_{\rm i}(u)<0$. In the interior region
of a BTZ black hole, we see from (\ref{ww}) that the constant $u$ null-geodesics reach the 
conical singularity at $r=0$ in a finite time given by $(1/\ell) \ln (-1/D_{\rm i}(u))$.
In contrast with the $n>3$ cases, the solutions $r(v)$ here do not reach $r=0$ perpendicularly (See Fig.
\ref{btz}). For the exterior region, the null-geodesics can   reach the spatial
infinity $r=\infty$ also in a finite time given by $(1/\ell) \ln (1/D_{\rm i}(u))$.

The solutions $r(v)$ for the mass function (\ref{ac3}) are obtained by matching, at $v=\pm v_0$,  
the BTZ black hole (\ref{ww}) with  the accretion process corresponding to $-v_0 <v < v_0$. This situation
is depicted in the Fig. \ref{btz}.

\section{Conclusion}

We considered here the problem of constructing the $n$-dimensional Vaidya metric
with a cosmological constant $\Lambda$ in double-null 
coordinates, generalizing the work \cite{GS}, where a semi-analytical approach
is proposed to
attack the equations derived previously by Waugh and Lake\cite{WL} for $n=4$ and $\Lambda=0$.
Some new exact solutions are also presented. For $n=3$, in particular, the problem
reduces to the solution of the second order linear ordinary differential equation
(\ref{ode}), allowing the analytical study of large classes of mass functions $A(v)$.
As an example, for $\Lambda <0$ and $A(v)= m_0+m_1\tanh v$, $m_0>m_1$, Eq. (\ref{ode}) is 
equivalent to the
hypergeometric equation, after a proper redefinition of $v$. This situation corresponds
to a BTZ black hole with mass $m_0-m_1$ that accretes some mass smoothly and ends
with mass $m_0+m_1$. This kind of smooth exact solution, with two ``asymptotic BTZ" regions
($v\rightarrow\pm\infty$) is very interesting to the study of creation of particles
in non stationary spacetimes\cite{Asympt}.

The generalized semi-analytical approach is also useful to the study of quasinormal
modes of varying mass $n$-dimensional black holes. In particular, an interesting
point is the study of the stationary regime for quasinormal modes, described in \cite{ACS},
in higher dimensional space and in the presence of a cosmological constant. The
semi-analytical approach can help the understanding of the relation anticipated by
Konoplya\cite{Konoplya} for the quasinormal modes of $n$-dimensional black
holes with $\Lambda=0$: $\omega_{R} \propto nr_0^{-1}$, where  $\omega_{R}$ and $r_0$ stand
for the oscillation
frequency and the black hole radius, respectively. Some preliminary results\cite{ACS2} suggests  
the existence of a stationary regime such that, for slowly varying mass $n$-dimensional
black holes, $\omega_{R}(v) \propto nr_0^{-1}(v)$.
As another application for $\Lambda\ne 0$, we have the dynamical formation of a Nariai solution,
which could be accomplished by choosing $A(v)$ such that $r_{EH}(v)\rightarrow r_{CH}(v)$ for larges
$v$. These points are still under investigation.

\acknowledgments

This work was supported by FAPESP and CNPq.

\appendix

\section{Some geometrical quantities}

We present here some geometrical quantities   necessary to
reproduce the equations of Section II. 
Such quantities are the $n$-dimensional $(n>2)$ generalization of those
ones calculated previously by Synge\cite{Synge} and Waugh and Lake\cite{WL} for $n=4$.
Only nonvanishing terms are given.

Recall that our $n$-dimensional line element is given by
\beq
ds^2 = -2f(u,v)du\,dv + r^2(u,v)d\Omega^2_{n-2},
\eeq
where $d\Omega^2_{n-2}$ is the unity $(n-2)$ 
sphere, assumed here to be spanned by the angular coordinates $ \theta_i,\ i=1\dots (n-2)$,
\beq
d\Omega^2_{n-2} = \sum_{i=1}^{n-2}\left(\prod_{j=1}^{i-1} \sin^2\theta_j \right) d\theta_{i}^2.
\eeq
The Christoffel symbols are
\begin{eqnarray}
\Gamma_{uu}^u &=& \frac{f_1}{f}, \quad \Gamma_{vv}^v = \frac{f_2}{f},  \\
\Gamma_{u\theta_k}^{\theta_k}  &=&  \frac{r_1}{r}, \quad \Gamma_{v\theta_k}^{\theta_k}  = \frac{r_2}{r} \nonumber\\
\Gamma_{\theta_k\!\theta_k}^u &=& r\left(\prod_{i=1}^{k-1}\sin^2\theta_i \right)\frac{r_2}{f}  \nonumber \\
\Gamma_{\theta_k\!\theta_k}^v &=& r\left(\prod_{i=1}^{k-1}\sin^2\theta_i \right)\frac{r_1}{f} \nonumber \\
\Gamma_{\theta_{k}\!\theta_{k}}^{\theta_j} &=& -\frac{1}{2}\left(\prod_{i=j+1}^{k-1} \sin^2\theta_i\right)\sin 2\theta_j,\quad (k>j)\nonumber \\
\Gamma_{\theta_j\!\theta_k}^{\theta_k} &=&   \cot  \theta_j, \quad (k\ne j)\nonumber 
\end{eqnarray}

The totally covariant Riemann tensor $R_{abcd}$ is given by
\begin{eqnarray}
\label{riemann}
R_{uvuv} &=& \frac{f_1f_2}{f} - f_{12} ,  \\
R_{u\theta_k\!u\theta_k}  &=&  r\left(\prod_{i=1}^{k-1}\sin^2\theta_i \right) \left( \frac{f_1r_1}{f} - r_{11}\right)  \nonumber \\
R_{v\theta_k\!v\theta_k} &=& r\left(\prod_{i=1}^{k-1}\sin^2\theta_i \right)  \left( \frac{f_2r_2}{f} - r_{22}\right), \nonumber \\
R_{u\theta_k\!\theta_k v}  &=&  r\left(\prod_{i=1}^{k-1}\sin^2\theta_i \right)    r_{12}   \nonumber  \\
R_{\theta_j\!\theta_k\!\theta_j\!\theta_k} &=& r^2
\left(\prod_{l=1}^{j-1} \sin^2 \theta_l \right)
\left(\prod_{i=j+1}^{k-1} \sin^2 \theta_i \right)  \left( 1 + 2\frac{r_1r_2}{f}\right) ,\quad (k>j). \nonumber
\end{eqnarray} 
Using (\ref{B}), we get from the purely angular components of the Riemann tensor:
\beq
\label{m}
R_{\theta_1\!\theta_k\!\theta_1 }^{\phantom{\theta_j\!\theta_k\!\theta_j\!}\theta_k} =   1 + \frac{r_2}{B}  ,
\eeq
for any $k>1$.
As for the $n=4$ case\cite{mass}, such components   are invariant under
transformations involving solely   the double-null coordinates $(u,v)\rightarrow (U(u,v),V(u,v))$ and
are candidates for defining the mass of the solution for $\Lambda=0$.

The 
Ricci tensor, obtained by the contraction $R_{ab} = R_{cab}^{\phantom{cab}c}$ is given by
\begin{eqnarray}
\label{ricci}
R_{uu} &=& (n-2) \left(\frac{f_1r_1}{fr} - \frac{r_{11}}{r} \right)  \\
R_{uv} &=&   \frac{f_1f_2}{f^2} - \frac{f_{12}}{f} - (n-2)\frac{r_{12}}{r} \nonumber \\
R_{vv} &=& (n-2) \left(\frac{f_2r_2}{fr} - \frac{r_{22}}{r} \right) \nonumber \\
R_{\theta_k\!\theta_k} &=& \left(\prod_{i=1}^{k-1} \sin^2\theta_i\right)\left[\frac{2}{f} \left(rr_{12} + (n-3)r_1r_2 \right) + (n-3)\right]\nonumber 
\end{eqnarray}

Two scalar quantities are relevant for our purposes: the scalar curvature $R=g^{ab}R_{ab}$ and the Kretschmann scalar 
$K=R_{abcd}R^{abcd}$. They are given by
\beq
R = \frac{2}{f}\left( f_{12} - \frac{f_1f_2}{f}\right)    + \frac{n-2}{r^2}\left( 
4\frac{r_1r_2}{f} + (n-3)\left( 1 + \frac{rr_{12}}{f}\right) 
\right), \nonumber
\eeq
and
\begin{widetext}
\begin{eqnarray}
K &=& 2\frac{(n-1)(n-3)}{r^2}
\left[ 
\frac{4}{f^2} \left( 
 r_{12}^2 + r_{11}r_{22} +\frac{r_1r_2f_1f_2}{f^2} - \frac{r_{22}r_1f_1}{f} - \frac{r_{11}r_2f_2}{f} 
\right) \right. \nonumber \\
&& \left. +\frac{1}{r^2} \left( 1 + 4\frac{r_1^2r_2^2}{f^2} +\frac{r_1r_2}{f} \right)
\right] 
+\frac{4}{f^2}\left( \frac{f_1^2f_2^2}{f^2} +f_{12}^2 - 2\frac{f_{12}f_1f_2}{f}\right).
\label{K}
\end{eqnarray}
\end{widetext}

\section{Some polynomial results}
\label{hor}

The evaluation of the zeros of $f$ given by (\ref{ff}) involves
finding the positive roots of the polynomial
\beq
\label{p1}
p_n(x) = x^{n-3} \left(x-\frac{1}{2} \right) = -\frac{\lambda}{n-3},
\eeq
for $\lambda > 0$ and $n>3$. The only roots of $p_n(x)$ are $x=1/2$ and $x=0$, the latter
with multiplicity $n-3$. For positive $x$, $p_n(x)<0$ only in the interval
$x\in (0,1/2)$. The minimal value of $p_n(x)$ in such an interval corresponds to
\beq
p_n(x^p_{\rm max}) = - \frac{1}{n-3}\lambda^c_{n} = - \frac{1}{n-3}\left( \frac{n-3}{2(n-2)}\right)^{n-2},
\eeq
where 
\beq
x_{\rm max}^p = \frac{n-3}{2(n-2)},
\eeq
the unique root of $p'_n(x)$ in the interval. 
Thus, for $\lambda > \lambda^c_n$, the polynomial (\ref{p1}) has no
roots. For $\lambda = \lambda^c_n$ there is only one root ($x=x^p_{\rm max}$), while for 
$0<\lambda<\lambda^c_n$, there are two roots $x_1$ and $x_2$, with $0<x_1<x^p_{\rm max}<x_2 < 1/2$.

The determination of 
 the curves (\ref{ahor})   involves the evaluation of the positive roots
of the polynomial 
\beq
\label{q1}
q_n(x)=x^{n-3}\left(1-\frac{2\Lambda}{(n-1)(n-2)}x^{2}\right) = \frac{2A}{n-3},
\eeq
with $A>0$ and $n>3$.  
For the  anti de Sitter case, $\Lambda <0$, the only root of $q_n(x)$ is $x=0$, $q_n(x)$ and $q'_n(x)$ are both positive for all $x > 0$, 
and, therefore, (\ref{q1})
   has only one positive root.
 Equation (\ref{ahor}), in this case, defines only one curve.

On the other hand, for the de Sitter case, $\Lambda >0$, $q_n(x)$ has always 3 roots: $x=\pm \sqrt{ (n-1)(n-2)/2\Lambda}$ and $x=0$, the
last with multiplicity $n-3$. Moreover, for positive $x$,   $q_n(x)\ge 0$ only 
for $x\in [0, \sqrt{ (n-1)(n-2)/2\Lambda}]$, and
the maximum value of $q_n(x)$ in such an interval is $q_n(x^q_{\rm max})$,
where
\beq
x_{\rm max}^q = \sqrt{\frac{(n-2)(n-3)}{2\Lambda}},
\eeq
the
only point of the interval where $q'_n(x)=0$. 
For $0 < {2A}/(n-3) < q_n(x_{\rm max}^q)$, 
(\ref{q1})  has always two roots $x_1$ and $x_2$, $x_1 < x_{\rm max}^q < x_2$. 
For ${2A}/(n-3) = q_n(x^q_{\rm max})$,
there is only one root $x=x_{\rm max}^q$, and for ${2A}/(n-3) > p_n(x^q_{\rm max})$ there 
is no root at all.
In this case, provided that $0 < {2A}/(n-3) < q_n(x^q_{\rm max})$, 
Eq. (\ref{ahor}) defines two curves $r_{EH}(v)$ and
$r_{CH}(v)$ (See Fig. \ref{figure1}), such that $r_{EH}(v) < x_{\rm max}^q < r_{CH}(v) $.

If $r(v)$ is a curve defined by (\ref{ahor}), its derivative is given by
\beq
r'(v) = \frac{2}{n-3}\frac{A'(v)}{r^{n-4}} \left((n-3) - \frac{2\Lambda}{n-2}r^2\right)^{-1}.
\eeq
If $\Lambda < 0$, it is clear that $r'(v)$ has the same sign of   $A'(v)$. 
For $\Lambda > 0$, the curves $r_{EH}(v)$ and $r_{CH}(v)$ have qualitatively
distinct behavior. Since $r_{EH}(v)<x_{\rm max}^q$, $r'_{EH}(v)$
has the same sign of $A'(v)$. The curve $r_{CH}(v)$  has the converse behavior,
since $r_{CH}(v)>x_{\rm max}^q$.

\end{document}